\documentstyle[12pt]{article}

\begin{document}

\setcounter{page}{1}

\pagestyle{plain} \vspace{1cm}
\begin{center}
\Large{\bf On the Stability of Planetary Circular Orbits in
Noncommutative Spaces}\\
\small \vspace{1cm}
\vspace{1cm} {\bf Kourosh Nozari} \quad and \quad {\bf Siamak Akhshabi}\\
\vspace{0.5cm} {\it Department of Physics,
Faculty of Basic Science,\\
University of Mazandaran,\\
P. O. Box 47416-1467,
Babolsar, IRAN\\
e-mail: knozari@umz.ac.ir}
\end{center}
\vspace{1.5cm}
\begin{abstract}
We investigate the effects of space noncommutativity and the
generalized uncertainty principle on the stability of circular
orbits of particles in both a central force potential and
Schwarzschild spacetime. We find noncommutative form of the
effective potential which up to first order of noncommutativity
parameter contains an angular momentum dependent extra term. This
angular momentum dependent extra term affects the stability of
circular orbits in such a way that the radius of a stable circular
orbit in noncommutative space is larger than its commutative
counterpart. In the case of large angular momentum, the condition
for stability of circular orbits in noncommutative space differs
considerably from commutative case.
\end{abstract}
\newpage
\section{ Introduction}
An important consequence of quantum gravity scenarios such as string
theory is the possible noncommutativity of spacetime structure at
short distances[1-7]. This noncommutativity leads to the
modification of Heisenberg uncertainty relations in such a way that
prevents one from measuring positions to better accuracies than the
Planck length. In low energy limit, these quantum gravity effects
can be neglected, but in circumstances such as very early universe
or in the strong gravitational field of a black hole one has to
consider these effects. The modifications induced by the generalized
uncertainty principle on the classical orbits of particles in a
central force potential firstly has been considered by Benczik {\it
et al}[8]. The same problem has been considered within
noncommutative geometry by Mirza and Dehghani[9]. The main
consequence of these two investigation is the constraint imposed on
the minimal observable length and noncommutativity parameter in
comparison with observational data of Mercury. Here we are going to
proceed one more step in this direction. We study the effects of the
generalized uncertainty principle and space noncommutativity on the
stability of circular orbits of particles in a central force
potential and also Schwarzschild geometry. We obtain a
noncommutative effective potential in each case, where up to first
order of noncommutativity parameter, there is an extra angular
momentum dependent term which affects the conditions for stability
of circular orbits. For large values of angular momentum, the effect
of space noncommutativity is considerable.

\section{Preliminaries}
In usual quantum mechanics one has the following canonical
commutation relations:
\begin{equation}
[\tilde{x_{i}},\tilde{x_{j}}]=0,   \,\,\,\,\,\,\,\,\,
[\tilde{p_{i}},\tilde{p_{j}}]=0,
   \,\,\,\,\,\,\,\,\,   [\tilde{x_{i}},\tilde{p_{j}}]=i\hbar\delta_{ij}
\end{equation}
From a string theory point of view, existence of a minimal length
scale can be addressed in the following generalized uncertainty
principle[10,11]
\begin{equation}
\delta{\tilde x_{i}}\delta {\tilde p_{i}} \geq \frac{\hbar}{2}
\bigg(1+\beta (\delta {\tilde p_{i}})^2+\gamma\bigg),
\end{equation}
where $\beta$ is string theory parameter related to minimal
length[11]. Since we are going to deal with absolutely minimum
position uncertainty, we set $\;\gamma=\beta { \langle {\tilde
p}\rangle^{2}}$ and therefore the corresponding canonical
commutation relation becomes
\begin{equation}
[\tilde {x}_{i},\tilde {p}_{j}]=i\hbar\delta_{ij}(1+\beta \tilde
p^{2}),
\end{equation}
where $\tilde p^{2}$ is the squared value of momentum. Now,
canonical commutation relations in commutative space with GUP
becomes
\begin{equation}
\left[ \tilde {x}_i,\tilde{ x}_j \right]=0,\;\;\;\,\, \left[ \tilde{
x}_i,\tilde{ p}_j \right]=i\hbar \delta_{ij}(1+\beta { \tilde p^{2}}
),\;\;\;\,\, \left[ \tilde{ p}_i,\tilde{ p}_j \right]=0.
\end{equation}
In a noncommutative space, the set of relations (1) should be
changed to the following relations
\begin{equation}
[\tilde{x_{i}},\tilde{x_{j}}]=i\theta_{ij},\quad\quad
[\tilde{p_{i}},\tilde{p_{j}}]=0,\quad\quad
[\tilde{x_{i}},\tilde{p_{j}}]=i\hbar\delta_{ij}
\end{equation}
where $\theta_{ij}$  is an anti-symmetric matrix which determines
the fundamental cell discretization of space-time in the same way as
the Planck constant $\hbar$  discretizes the phase space. The
elements of $\theta_{ij}$ have the dimension of $(length)^{2}$  . In
the classical limit, the commutators should be replaced by
corresponding Poisson brackets
\begin{equation}
\frac{1}{i\hbar}[\tilde{A},\tilde{B}]\longrightarrow\{\tilde{A},\tilde{B}\}
\end{equation}
and therefore in this limit the commutation relations become as
follows
\begin{equation}
\{\tilde{x_{i}},\tilde{x_{j}}\}=\alpha_{ij},\quad\quad
\{\tilde{p_{i}},\tilde{p_{j}}\}=0,\quad\quad
\{\tilde{x_{i}},\tilde{p_{j}}\}=\delta_{ij}
\end{equation}
here $\alpha_{ij}=\frac{\Theta_{ij}}{\hbar} $ . The Poisson bracket
should possess the same properties as the quantum mechanical
commutator, namely, it should be bilinear , anti-symmetric and
should satisfy the Leibniz rules and the Jacobi identity . The
general form of the Poisson brackets for this deformed  version of
classical mechanics can be written as follows [8,9] :
\begin{equation}
\{A,B\}=\Bigg[\frac{\partial
A}{\partial{\tilde{x_{i}}}}\frac{\partial
B}{\partial\tilde{p_{j}}}-\frac{\partial A}{\partial
\tilde{p_{i}}}\frac{\partial
B}{\partial\tilde{x_{j}}}\Bigg]\{\tilde{x_{i}},\tilde{p_{j}}\}+\frac{\partial
A}{\partial\tilde{x_{i}}}\frac{\partial
B}{\partial\tilde{x_{j}}}\{\tilde{x_{i}},\tilde{x_{j}}\}
\end{equation}
where repeated indices are summed. There is a new coordinate system
which can be defined with the following transformations
\begin{equation}
x_{i}=\tilde{x_{i}}+\frac{1}{2}\alpha_{ij}\tilde{p_{j}}, \quad\quad
p_{i}=\tilde{p_{i}}
\end{equation}
where the new variables satisfy the usual canonical brackets
\begin{equation}
\{x_{i},x_{j}\}=0, \quad\quad \{p_{i},p_{k}\}=0,\quad\quad
\{x_{i},p_{j}\}=\delta_{ij}.
\end{equation}
With these preliminaries, we have all prerequisites to investigate
particle orbits and their stability within noncommutative geometry
and the generalized uncertainty principle.

\section{ Central Force Potential}
In noncommutative space, for a particle in a central force potential
the Hamiltonian has the following form
\begin{equation}
H=\frac{\tilde{p}^{2}}{2m}+V(\tilde{r}),\quad\quad \tilde
{r}=\sqrt{\tilde{x_{i}}\tilde{x_{j}}}
\end{equation}
Using coordinates transformation given as (9), we find
$$
V(\tilde{r})=V\Bigg(\sqrt{\big(x_{i}-\frac{\alpha_{ij}p_{j}}{2}\big)
\big(x_{i}-\frac{\alpha_{ik}p_{k}}{2}\big)}\Bigg)$$ $$
=V(r)+\frac{\vec{\alpha}\times\vec{p}}{2}\cdot\vec{\nabla}
V(r)+O(\alpha^{2})$$
\begin{equation}
 =V(r)-\frac{(\vec{\alpha}\cdot\vec{L})}{2r}\frac{\partial V}{\partial
 r}+O(\alpha^{2}),
\end{equation}
where $\alpha_{ij}=\epsilon_{ijk}\alpha_{k}$ and
$\vec{L}=\vec{r}\times\vec{p} $. In which follows, we first
determine the effects of space noncommutativity on the planetary
orbits and then we investigate the stability of circular orbits in a
general central force potential.

\subsection{Particle Orbits}
The effect of space noncommutativity on orbital motions of particles
in a central force potential has been investigated by Mirza and
Dehghani[9]. For an inverse squared force(Coulomb force) as
$F=-\frac{k}{r^{2}}$, the Hamiltonian up to the first order of
$\alpha$ becomes
\begin{equation}
H=\frac{p^{2}}{2m}-\frac{k}{r}-\frac{k}{r^{3}}\Big(\frac{\vec{\alpha}
\cdot\vec{L}}{2}\Big).
\end{equation}
The new term in the Hamiltonian is small and its effects can be
obtained by standard perturbation theory, however it causes a time
dependent angular momentum. We assume that the time variation of the
$\vec L $ is so small that in a short time interval (for example one
century for Mercury) it could be taken as a constant of the motion.
In this way, one can put a bound on the value of $\alpha $ by
comparing the results of the perturbing term with the experimental
value of the precession of the perihelion of Mercury[8,9]. For the
Kepler problem it is well known that the bounded orbits are closed,
which is a result of the following integral,
\begin{equation}
\triangle \varphi^{(0)}  = -2 {\partial \over
\partial L}\int_{r_{min}}^{r_{max}} { \sqrt {2 m \Big[E- V(r)\Big] - {L^2 \over r^2
}}}dr = 2 \pi,
\end{equation}
where
$$ E= {1\over 2} m(\dot r^2 + r^2 \dot {\varphi}^2) + V(r) ,\quad\quad
L=m r^2 \dot {\varphi}.$$ By perturbing the potential with a small
term $ V \rightarrow V + \delta V $ and expanding the integral up to
the first order in $ \delta V $, we find
\begin{equation}
\triangle \varphi = \triangle \varphi^{(0)} + \triangle
\varphi^{(1)}
\end{equation}
where
\begin{equation}
\triangle \varphi^{(1)}= {\partial \over \partial L}{\int_0^\pi
\Big({{2 m r^2}\over L}\Big)\  \delta V \  d\varphi }.
\end{equation}
After a straightforward calculation, we arrive at the following
relation
\begin{equation}
\triangle \varphi^{(1)}=\Big( { {2  \pi k^2  m^2   cos( \psi )
}\over L^3}\Big)\alpha
\end{equation}
where $\psi $ is the angle between $\vec \alpha $ and $\vec L $.
This is the effect of space noncommutativity on the orbits of
particles in a central force potential. Comparison with
observational data of Mercury sets a bound on the noncommutativity
parameter[9].

\subsection{Stability of the Circular Orbits}
Now we consider a central force with the following form
\begin{equation}
F=-\frac{k}{r^{n}}.
\end{equation}
In this case, the effective potential up to first order of $\alpha$
can be written as
\begin{equation}
V_{eff}(\tilde{r})=-\frac{k}{n-1}\frac{1}{r^{n-1}}+\frac{L^{2}}{2mr^{2}}-
\frac{k}{r^{n+1}}(\frac{\vec{\alpha}\cdot\vec{L}}{2}).
\end{equation}
Here $m $ is the  mass of the particle and $L=mr^{2}\dot{\varphi}$.
The particle could have a stable circular orbit at  $r=\rho$  if the
effective potential has a real minimum at $\rho $. In other words,
the effective potential should satisfy the following conditions to
have stable circular orbit at $\rho$
\begin{equation}
\frac{\partial V_{eff}}{\partial r}|_{r=\rho}=0,\quad\quad and
\quad\quad \frac{\partial^{2}V_{eff}}{\partial r^{2}}|_{r=\rho}>0.
\end{equation}
With effective potential as (19), we find
\begin{equation}
\frac{\partial V_{eff}}{\partial r}|_{r=\rho}=\frac{k}{\rho^{n}}-
\frac{L^{2}}{m\rho^{3}}+
\Big(\frac{\vec{\alpha}\cdot\vec{L}}{2}\Big)\frac{k(n+1)}{\rho^{n+2}}=0,
\end{equation}
and
\begin{equation}
\frac{\partial^{2}V_{eff}}{\partial
r^{2}}|_{r=\rho}=-\frac{nk}{\rho^{n+1}}+
\frac{3L^{2}}{m\rho^{4}}-\Big(\frac{\vec{\alpha}\cdot\vec{L}}{2}\Big)\frac{k(n+1)(n+2)}
{\rho^{n+3}}>0.
\end{equation}
These two relations can be rewritten as follows respectively
\begin{equation}
k\rho^{2}-\frac{L^{2}}{m}\rho^{n-1}+\Big(\frac{\vec{\alpha}.\vec{L}}{2}\Big)k(n+1)=0,
\end{equation}
and
\begin{equation}
-nk\rho^{2}+\frac{3L^{2}}{m}\rho^{n-1}-(\frac{\vec{\alpha}\cdot\vec{L}}{2})
k(n+1)(n+2)>0.
\end{equation}
From equation (23) we find
\begin{equation}
\frac{L^{2}}{m}\rho^{n-1}=k\rho^{2}+(\frac{\vec{\alpha}.\vec{L}}{2})k(n+1).
\end{equation}
Using this result in (24), we obtain the following condition for
stability of circular orbits in central force of the type given as
(18)
\begin{equation}
(3-n)k\rho^{2}-\Big(\frac{\vec{\alpha}\cdot\vec{L}}{2}\Big)k(n+1)(n-1)>0,
\end{equation}
or
\begin{equation}
(3-n)k\rho^{2}-\frac{k(n+1)(n-1)L\alpha\cos(\psi)}{2}>0,
\end{equation}
where $\psi$  is the angle between  $ \vec{\alpha}$ and $\vec{L} $.
In the absence of space noncommutativity , i.e., when $\alpha=0 $,
we find the usual result of $n<3 $, as it is well known from
classical mechanics. In the presence of space noncommutativity, the
third term in effective potential (19) causes a time dependent
angular momentum which affects the stability of circular orbits.
However the time variation of $\vec{L}$ is so small that in a short
time interval it could be taken as a constant of the motion. Figure
1 shows the conditions for stability of circular orbits in a central
force of the type given in (18). As this figures shows, for small
values of $L$, the condition for stability of circular orbits is the
same as classical commutative result, i. e. , $n<3$. When $L$
increases, the condition for stability differs considerably relative
to commutative case. In this case $n$ can attains non-integer
values. This feature may be interpreted as the possible deviation
from inverse square law. In another words, possible deviation from
inverse square law can be interpreted as a consequence of space
noncommutativity.

\subsection{Stability of Circular Orbits in Yukawa Potential}
As another specific example, here we consider the Yukawa potential
and investigate the stability of its circular orbits in
noncommutative space. Yukawa potential is defined as follows
\begin{equation}
V(r)=-\frac{k}{r}e^{-r/a} \quad\quad\quad\quad      k>0,\quad  a>0.
\end{equation}
Using equation (12), the corresponding potential in noncommutative
space becomes
\begin{equation}
V(\tilde{r})=-\frac{k}{r}e^{-r/a}-k\Big(\frac{\vec{\alpha}\cdot\vec{L}}{2}\Big)
\Big(\frac{1}{ar^{2}}+\frac{1}{r^{3}}\Big)e^{-r/a},
\end{equation}
and therefore the effective potential is given by
\begin{equation}
V_{eff}(\tilde{r})=-\frac{k}{r}e^{-r/a}-k\Big(\frac{\vec{\alpha}\cdot\vec{L}}{2}\Big)
\Big(\frac{1}{ar^{2}}+\frac{1}{r^{3}}\Big)e^{-r/a}+\frac{L^{2}}{2mr^{2}}.
\end{equation}
By applying the conditions (20) to this $ V_{eff}(\tilde{r})$, a
straightforward calculation gives the following condition for
stability of circular orbits in Yukawa potential
\begin{equation}
3+\rho\frac{F'(\rho)}{F(\rho)}>0,
\end{equation}
where $F(r)=-\frac{\partial V}{\partial r}$  and
$F'(r)=\frac{\partial F}{\partial r} $. A simple calculation gives
\begin{equation}
3-\frac{[\frac{2}{\rho^{2}}+\frac{2}{a\rho}+\frac{1}{\rho^{2}}]+[\frac{\vec{\alpha}
\cdot\vec{L}}{2}][\frac{12}{\rho^{4}}+\frac{12}{a\rho^{3}}+\frac{5}{a^{2}\rho^{2}}+
\frac{1}{a^{3}\rho}]}{[\frac{1}{a\rho}+\frac{1}{\rho^{2}}]+[\frac{\vec{\alpha}
\cdot\vec{L}}{2}][\frac{3}{\rho^{4}}+\frac{3}{a\rho^{3}}+\frac{1}{a^{2}\rho^{2}}]}>0,
\end{equation}
or
\begin{equation}
(a^{2}\rho^{3}+a^{3}\rho^{2}-a\rho^{4})-\Big(\frac{\vec{\alpha}\cdot\vec{L}}{2}\Big)
\Big(\rho^{3}+2a\rho^{2}+3a^{2}\rho+3a^{3}\Big)>0.
\end{equation}
This is the condition for stability of circular orbits of Yukawa
potential in noncommutative space. If $\alpha=0$, this relation
reduces to the following condition
\begin{equation}
a\rho+a^{2}-\rho^{2}=0,
\end{equation}
which leads to the condition
$$\rho\leq1.62a $$
for stability of circular orbits. This result is well known from
classical mechanics.\footnote{A digression to a newly proposed
scenario may be helpful here. Equation (34) has the following
solutions
$$\rho=\frac{a}{2}\pm\sqrt{\frac{a^2}{4}+1}=a\frac{1}{2}\Big[1\pm\sqrt{\frac{5}{4}}\Big]$$
where in the language of E-infinity theory[12,13] can be interpreted
as follows
$$\rho_{+}=a\frac{1}{2}[1+\sqrt{5}]=a(\frac{1}{\phi})$$
and
$$\rho_{-}=a\frac{1}{2}[1-\sqrt{5}]=-a(\phi)$$
where $\phi$ is the golden mean. Note also that in this framework
$$\rho\leq (1.618033969) a.$$.}

Figure 2  shows the condition for stability of circular orbits of
particles in Yukawa potential. As this figure shows, for small
values of $L$, the condition for stability is essentially the same
as commutative case. For large $L$ however the situation is
different from commutative case. For large $L$, we see that the
condition can change to, for example, $$\rho\leq 1.414213562a. $$
Note that since noncommutativity parameter is very small, the role
played by $\vec{L}$ is enhanced. But we know that presence of $\vec
{L}$ is due to non-vanishing value of $\alpha$, that is, space
noncommutativity. Thus, we can conclude that space noncommutativity
(up to first order of noncommutativity parameter) causes
$\vec{L}$-dependent terms in orbital equations which these terms are
important only in large values of $L$.

\section{Stability of Circular Orbits in Schwarzschild Geometry}
In this section we investigate the existence and stability of
circular orbits of particles in noncommutative Schwarzschild
geometry. We first obtain commutative effective potential and then
generalize it to noncommutative case.\\
Consider the geometry of Schwarzschild spacetime with the following
metric(with $c=1$)
\begin{equation}
ds^{2}=-(1-\frac{2GM}{r})dt^{2}+(1-\frac{2GM}{r})^{-1}dr^{2}+r^{2}d\Omega^{2}
\end{equation}
The geodesics of this spacetime are given by the following
equations[14]
\begin{equation}
\frac{d^{2}t}{d\lambda^{2}}+\frac{2GM}{r(r-2GM)}\frac{dr}{d\lambda}\frac{dt}{d\lambda}
=0
\end{equation}
$$\frac{d^{2}r}{d\lambda^{2}}+\frac{GM}{r^{3}}(r-2GM)(\frac{dt}
{d\lambda})^{2}-\frac{GM}{r(r-2GM)}(\frac{dr}{d\lambda})^{2}$$
\begin{equation}
-(r-2GM) \Big[(\frac{d\vartheta}{d\lambda})^{2}+
\sin^{2}(\vartheta)(\frac{d\varphi}{d\lambda})^{2}\Big] =0
\end{equation}
\begin{equation}
\frac{d^{2}\vartheta}{d\lambda^{2}}+\frac{2}{r}\frac{d\vartheta}
{d\lambda}\frac{dr}{d\lambda}-\sin(\vartheta)
\cos(\vartheta)(\frac{d\varphi}{d\lambda})^{2}
=0
\end{equation}
\begin{equation}
\frac{d^{2}\varphi}{d\lambda^{2}}+\frac{2}{r}\frac{d\varphi}
{d\lambda}\frac{dr}{d\lambda}+2\cot(\vartheta)\frac{d\vartheta}{d\lambda}\frac{d\varphi}
{d\lambda}=0
\end{equation}
where $\lambda$ is an affine parameter. There are four conserved
quantities associated with Killing vectors: magnitude of angular
momentum(one component), direction of angular momentum(two
components) and energy. The two Killing vectors which lead to
conservation of the direction of angular momentum imply that
$\vartheta=\frac{\pi}{2}$. The other two Killing vectors are
corresponding to energy and the magnitude of the angular momentum.
The Killing vector associated with energy is $\partial_{t}$ or
\begin{equation}
K_{\mu}=\Big(-(1-\frac{2GM}{r}),\,\,0,\,\,0,\,\,0\Big)
\end{equation}
and for the magnitude of angular momentum the Killing vector is
$\partial_{\varphi}$ or
\begin{equation}
L_{\mu}=\Big(0,\,\,0,\,\,0,\,\,r^{2}\sin^{2}(\vartheta)\Big).
\end{equation}
So, along the geodesics the two corresponding conserved quantities
are
\begin{equation}
\Big(1-\frac{2GM}{r}\Big)\frac{dt}{d\lambda}=E
\end{equation}
and
\begin{equation}
r^{2}\frac{d\varphi}{d\lambda}=L
\end{equation}
respectively, where $E$ and $L$ are energy and angular momentum of
the particle per its unit mass.

Along an affinely parameterized geodesic ( timelike, spacelike or
null), the scalar quantity $$ \varepsilon=-u^{\alpha}u_{\alpha}$$ is
a constant since
$$\frac{d\varepsilon}{d\lambda}=(u^{\alpha}u_{\alpha})_{;\beta}u^{\beta}
=(u_{;\beta}^{\alpha}u^{\beta})u_{\alpha}+u^{\alpha}(u_{\alpha;\beta}u^{\beta})=0.$$
If we chose $\lambda$ to be proper time or proper distance,  then we
find $\varepsilon=\pm 1$. For a null geodesic $ \varepsilon=0$.
Generally, one can write
\begin{equation}
\varepsilon=-g_{\mu\nu}\frac{dx^{\mu}}{d\lambda}\frac{dx^{\nu}}{d\lambda}
\end{equation}
We chose $\lambda=\tau$ (the proper time) and expand this equation.
In equatorial plane ($\vartheta=\frac{\pi}{2})$  we find
\begin{equation}
-(1-\frac{2GM}{r})(\frac{dt}{d\lambda})^{2}+(1-\frac{2GM}{r})^{-1}(\frac{dr}
{d\lambda})^{2}+r^{2}(\frac{d\varphi}{d\lambda})^{2}=-\varepsilon
\end{equation}
where multiplying by $(1-\frac{2GM}{r})$ and using equations (48)
and (49), we obtain
\begin{equation}
-E^{2}+(\frac{dr}{d\lambda})^{2}+(1-\frac{2GM}{r})(\frac{L^{2}}{r^{2}}+\varepsilon)=0
\end{equation}
This relation can be rewritten as follows
\begin{equation}
\frac{1}{2}(\frac{dr}{d\lambda})^{2}+V_{eff}(r)=\frac{1}{2}E^{2}
\end{equation}
where we have defined
\begin{equation}
V_{eff}(r)=\frac{1}{2}\varepsilon-\varepsilon\frac{GM}{r}+\frac{L^{2}}{2r^{2}}-\frac{GML^{2}}{r^{3}}
\end{equation}
which is the {\it effective potential} in commutative Schwarzschild
spacetime. Figure $3$ shows the variation of this effective
potential versus radius  and for different angular momentum. In
terms of the Schwarzschild radius, one can rewrite this equation as
follows
\begin{equation}
V_{eff}(r)=\frac{1}{2}\varepsilon-\varepsilon\frac{GM}{r}+\frac{L^{2}}{2r^{2}}
(1-\frac{r_{s}}{r})
\end{equation}
where $ r_{s}$ is the Schwarzschild radius. We use this relation in
our forthcoming arguments.

In classical general relativity, the particle could have a circular
orbit at $r_{c}$ if
$$\Bigg(\frac{dV}{dr}\Bigg)_{r=r_c}=0 $$
which leads to the following condition
\begin{equation}
\varepsilon GMr^{2}_{c}-L^{2}r_{c}+3GML^{2}\gamma=0
\end{equation}
where $ \gamma=0$ in Newtonian regime and $ \gamma=1$ in general
relativity[14]. Let us consider two possible cases. Firstly, for $
\gamma=1$ and $\varepsilon=0$ ( photons ) we have
\begin{equation}
r_{c}=3GM
\end{equation}
Secondly, for  $ \gamma=1$ and $\varepsilon=1$ ( massive particles )
we find
\begin{equation}
r_{c}=\frac{L^{2}\pm\sqrt{L^{4}-12G^{2}M^{2}L^{2}}}{2GM}
\end{equation}
For $ L\rightarrow\infty$, there exist a stable circular orbit at $
\frac{L^{2}}{GM}$ which goes farther and farther away and an
unstable one at $ 3GM$ . For small $ L$, at $ L=\sqrt{12}GM $ two
circular orbits coincide at $ r_{c}=6GM$ and disappear entirely for
smaller $L$. Therefore, $ 6GM$ is the smallest possible radius of a
stable circular orbit in Schwarzschild metric . In brief,
Schwarzschild solution in commutative space possesses stable
circular orbits for $ r\geq6GM $ and unstable ones for $ 3GM<r<6GM
$. Now we consider the effects of space noncommutativity on the
stability of circular orbits in Schwarzschild geometry.

\section{Noncommutative Space Considerations}
In Schwarzschild spacetime, with $\varepsilon=1$ ( the case of
timelike geodesics) we find from (48)
\begin{equation}
V(r)=\frac{1}{2}-\frac{GM}{r}+\frac{L^{2}}{2r^{2}}-\frac{GML^{2}}{r^{3}}.
\end{equation}
Considering the effects of space noncommutativity as given by (12),
this relation changes to the following form
\begin{equation}
V(\tilde{r})=\frac{1}{2}-\frac{GM}{r}+\frac{L^{2}}{2r^{2}}-
\frac{GML^{2}}{r^{3}}-\frac{L\theta\cos\psi}{2}\Big[\frac{GM}{r^{3}}-\frac{L^{2}}{r^{4}}+\frac{3GML^{2}}{r^{5}}\Big]
\end{equation}
where $\psi$ is the angle between $\vec{L}$ and $\vec {\theta}$.
This noncommutative effective potential has been plotted in figures
$4$ and $5$ for two different values of angular momentum. As these
figures show, increasing the value of the particle angular momentum
increases the difference between commutative and noncommutative
effective potentials. In another words, large angular momentum
enhances the effect of space noncommutativity. So space
noncommutativity couples with angular momentum. This point has its
origin on the algebraic structure of theory on noncommutativity
level.

The particle could have a circular orbit at $r_{c}$ if
$$\Bigg(\frac{dV}{dr}\Bigg)_{r=r_{c}}=0 $$ or
\begin{equation}
\frac{GM}{r^{2}_{c}}-\frac{L^{2}}{r^{3}_{c}}+\frac{3GML^{2}}{r^{4}_{c}}
-\frac{L\theta\cos\psi}{2}\Big[\frac{-3GM}{r^{4}_{c}}+\frac{4L^{2}}{r^{5}_{c}}-\frac{15GML^{2}}{r^{6}}\Big]=0\
\end{equation}
This can be simplified to find
\begin{equation}
GMr^{4}_{c}-L^{2}r_{c}^{3}+3GML^{2}r^{2}_{c}-\frac{L\theta\cos\psi}{2}\Big[-3GMr^{2}_{c}+4L^{2}r_{c}-15GML^{2}\Big]=0.
\end{equation}
This is the condition for existence of circular orbits in
noncommutative Schwarzschild geometry up to first order of
noncommutativity parameter. Comparing this equation with equation
(50)(with $\gamma=1$ and $\epsilon=+1$) for commutative case, shows
the importance of noncommutativity effect. Since noncommutativity
parameter is very small, this effect can be neglected in ordinary
circumstances but for large angular momentum the situation differs
considerably. The dependence of this modification to angular
momentum is a pure noncommutative effects which goes back to the
algebraic structure of the theory.

The condition for the stability of the circular orbits is
\begin{equation}
\Bigg(\frac{\partial^{2}V}{\partial r^{2}}\Bigg)_{r=r_c}>0
\end{equation}
Applying this condition to the potential (54), we find
\begin{equation}
\frac{-2GM}{r^{3}_{c}}+\frac{3L^{2}}{r^{4}_{c}}-\frac{12GML^{2}}{r^{5}_{c}}-
\frac{L\theta
\cos\psi}{2}\Big[\frac{12GM}{r^{5}_{c}}-\frac{20L^{2}}{r^{6}_{c}}+\frac{90GML^{2}}{r^{7}_{c}}\Big]>0
\end{equation}
Combining this equation with equation (55), we find the following
condition for the stability of the circular orbits in noncommutative
space
\begin{equation}
GMr^{4}_{c}-3GML^{2}r^{2}_{c}-\frac{L\theta
\cos(\psi)}{2}\Big[3GMr^{2}_{c}-8L^{2}r_{c}+45GML^{2}\Big]>0
\end{equation}
Figure $6$ shows the condition for stability of circular orbits of
particles in Schwarzschild geometry and in the presence of space
noncommutativity. In this figure we have plotted the left hand side
of relation (58) versus $r=GM$. As this figure shows, in the case of
noncommutative space, particles could have stable circular orbits
for $r\geq(6.139234690)GM$ which is greater than commutative result
of $r\geq6GM$. Therefore, stable circular orbits
have greater radius in noncommutative spaces.\\

\section{The effect of GUP}
Now we consider the effect of the generalized uncertainty principle
on the stability of circular orbits of particles in Schwarzschild
spacetime.  With the generalized uncertainty principle, the standard
commutation relations transform to the following general form [6,12]
\begin{equation}
[x_{i},p_{j}]=i\hbar(\delta_{ij}+\beta p^{2}\delta_{ij}+\beta'
p_{i}p_{j}),
\end{equation}
\begin{equation}
[p_{i},p_{j}]=0,
\end{equation}
and
\begin{equation}
[x_{i},x_{j}]=i\hbar\frac{(2\beta-\beta')+(2\beta+\beta')\beta
p^{2}}{(1+\beta p^{2})}(p_{i}x_{j}-p_{j}x_{i}).
\end{equation}
we set $\beta'=0$ so the corresponding Poisson brackets are
\begin{equation}
\{x_{i},p_{j}\}=\delta_{ij}(1+\beta p^{2}),\,\,\,\,\,\,\,
\{p_{i},p_{j}\}=0,\,\,\,\,\,\,\,\{x_{i},x_{j}\}=2\beta
(p_{i}x_{j}-p_{j}x_{i})
\end{equation}
The modified canonical equations now take the following forms
\begin{equation}
\dot{x_{i}}=\{x_{i},H\},\,\,\,\,\,\,\dot{p_{i}}=\{p_{i},H\}
\end{equation}
where
\begin{equation}
H=\frac{p^{2}}{2m}+V(r)
\end{equation}
is the Hamiltonian of the system. The deformed angular momentum
which is given by
\begin{equation}
L_{ij}=\frac{x_{i}p_{j}-x_{j}p_{i}}{(1+\beta p^{2})}
\end{equation}
is conserved due to the rotational symmetry of the  Hamiltonian. For
the circular orbits of particles in a general central force problem
we have the constraints $\dot{r}=0$\, and\, $\dot{p_{r}}=0$[8].
These conditions will led us to
\begin{equation}
L=\frac{pr}{(1+\beta p^{2})}.
\end{equation}
The radius of orbit $r$ and the magnitude of the momentum
$p=\sqrt{p^{2}}$\, are now the constants of the motion .

Now in Schwarzschild spacetime the effective potential is given by
equation (53). We substitute the value of $L$ from (67) in this
potential to find
\begin{equation}
V(\tilde{r})=\frac{1}{2}-\frac{GM}{r}+\frac{p^{2}}{2(1+\beta
p^{2})}-\frac{GMp^{2}}{r(1+\beta p^{2})}
\end{equation}
The radius of the circular orbits will be given by
\begin{equation}
r_{c}=1/2\,{\frac {GM \left(
2+4\,\beta\,{p}^{2}+2\,{\beta}^{2}{p}^{4}+3\, {p}^{2} \right)
}{{p}^{2}}}
\end{equation}
so the condition for a circular orbit with radius $r$ to be stable
in the framework of GUP is
\begin{equation}
-2GMr^{2}_{c}+\frac{3p^{2}r^{3}_{c}}{(1+\beta
p^{2})^{2}}-\frac{12GMp^{2}r^{3}_{c}}{(1+\beta p^{2})^{2}}\geq0
\end{equation}
We emphasize that presence of $\beta$ is a quantum gravitational
effect which has origin on the fractal structure of spacetime at
very short distances(string scale). The situation is much similar to
the case presented in figure 6 with dotted curve. The effect of
space noncommutativity is much similar to the effect of generalized
uncertainty principle. These two concept are common features of
quantum gravity era where there exist a minimal observable length of
the order of Planck length.

\section{Summary and Conclusion}
In this paper we have studied the effects of space noncommutativity
and the generalized uncertainty principle on the stability of
circular orbits of particles in both a central force potential and
Schwarzschild geometry. In the case of central force potential, we
have shown that when the angular momentum of the particle is small,
the effect of space noncommutativity can be neglected. In this case
for a central force of the type $F(r)=-\frac{k}{r^n}$, we find $n<3$
which is condition for stability of circular orbits in commutative
classical mechanics. The situation however differs considerably with
classical prescription in the case of large angular momentum. In
this case, the condition for stability of circular orbits differs
considerably with its classical commutative counterpart. Since the
effect of space noncommutativity in first order approximation is the
presence of angular momentum in orbital equations, we conclude that
space noncommutativity changes conditions for stability of the
circular orbits in an angular momentum dependent manner. For
clarifying our findings, we have considered the case of Yukawa
potential as an example.For the case $n=2$ in $F(r)=-\frac{k}{r^n}$,
from a more deeper point of view, all of our calculations show the
implicit departure from inverse squared law due to space
noncommutativity. This feature provides an operative approach for
investigation of the departure from inverse square law in
electricity and also planetary
physics.\\
 we have studied the effect of space noncommutativity and the
generalized uncertainty principle on the stability of circular
orbits of particles in Schwarzschild geometry. We have found the
effective potential in the case of noncommutative Schwarzschild
space. Again, when the angular momentum of the particle is small,
the effect of space noncommutativity can be neglected. In this case
we find $r\geq 6GM$ for the stability of circular orbits in
commutative Schwarzschild space. The situation differs considerably
with commutative prescription in the case of large angular momentum.
In this case the condition for stability of circular orbits for
$L=5$(in arbitrary units) becomes $r\geq(6.139234690)GM$ which
differs from commutative result, $r\geq6GM$. Since the effect of
space noncommutativity in first order approximation is the presence
of angular momentum in orbital equations, we conclude that space
noncommutativity changes conditions for stability of the circular
orbits in a angular momentum dependent manner. This feature provides
an operative approach for investigation of the spacetime geometry,
although this effects are so small that current experimental devices
can not detect them.

\begin{figure}[ht]
\begin{center}
\includegraphics{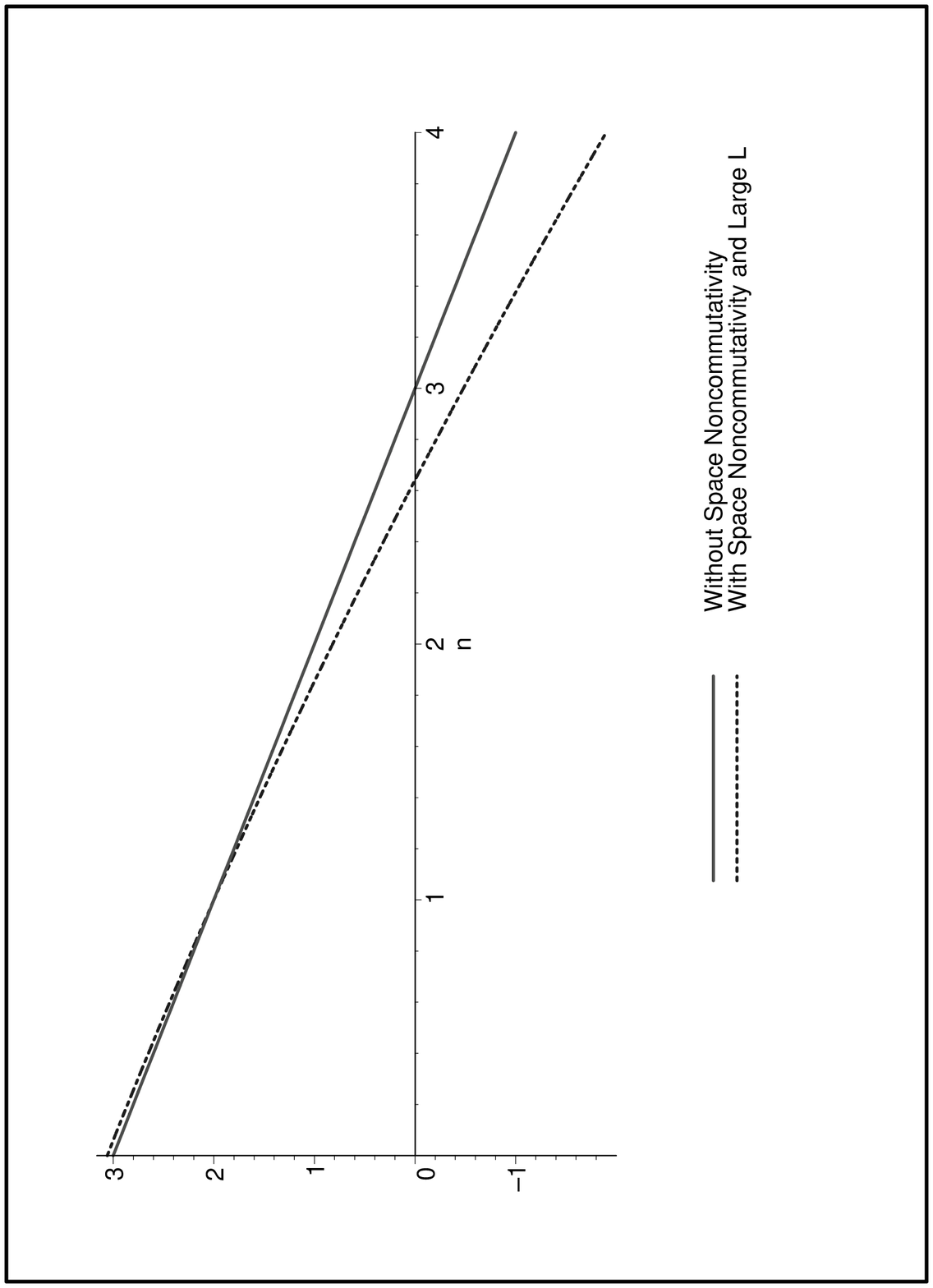}
\end{center}
\vspace{16 cm} \caption{\small {The effect of space noncommutativity
on the stability of circular orbits of particles in a central force
of the type $F(r)=-\frac{k}{r^n}$. Vertical axis shows the left hand
side of relation (29). Note that in $E$-infinity theory, based on El
Naschie bijection formula one can write
$D_{3}=\Big(\frac{1}{d_{c}^{(0)}}\Big)^{3-1}=
2+\phi=(\frac{1}{\phi})^{2}=2.618033969$ which is Hausdorff
dimension of Penrose universe when time dimension is added. On the
other hand this value is the intersection point of curve
corresponding to noncommutative case with horizontal axis. This can
be interpreted as $(\frac{1}{\phi})^{2}= 1+1.618033969$ where
$1.618033969$ is the golden mean[12,13].}}
 \label{fig:1}
\end{figure}

\begin{figure}[ht]
\begin{center}
\includegraphics{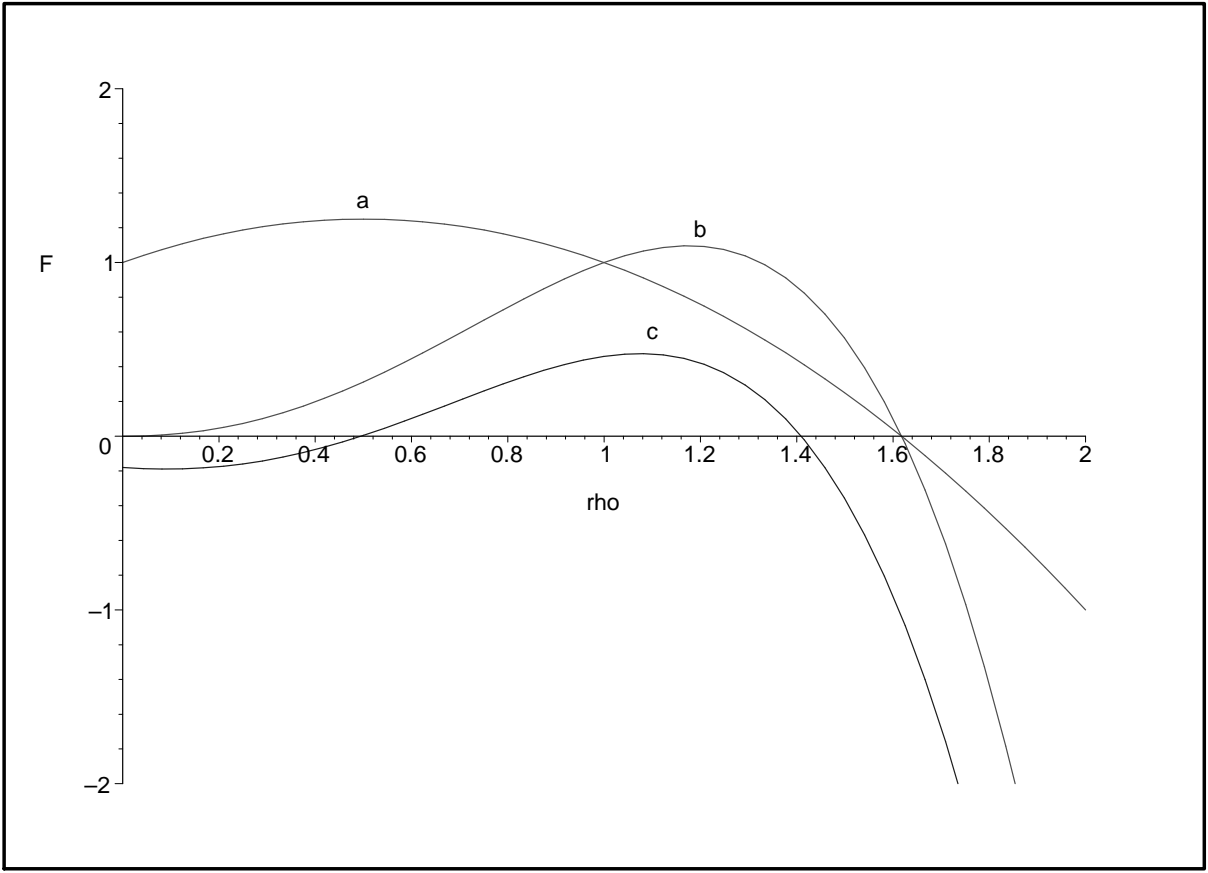}
\end{center}
\vspace{16 cm} \caption{\small {The effects of space
noncommutativity on the stability of circular orbits of particles in
Yukawa potential. a) Commutative space, b)Noncommutative space with
small L, and c)Noncommutative space with large L. Vertical axis
shows the left hand side of relation (35). In the language of
$E$-infinity theory, curves (a) and (b) intersect horizontal axis in
$1.618033969$ which is golden mean while curve (c) intersects
horizontal axis at silver mean, $1.414213562$ [12,13]. }}
 \label{fig:2}
\end{figure}

\begin{figure}[ht]
\begin{center}
\includegraphics{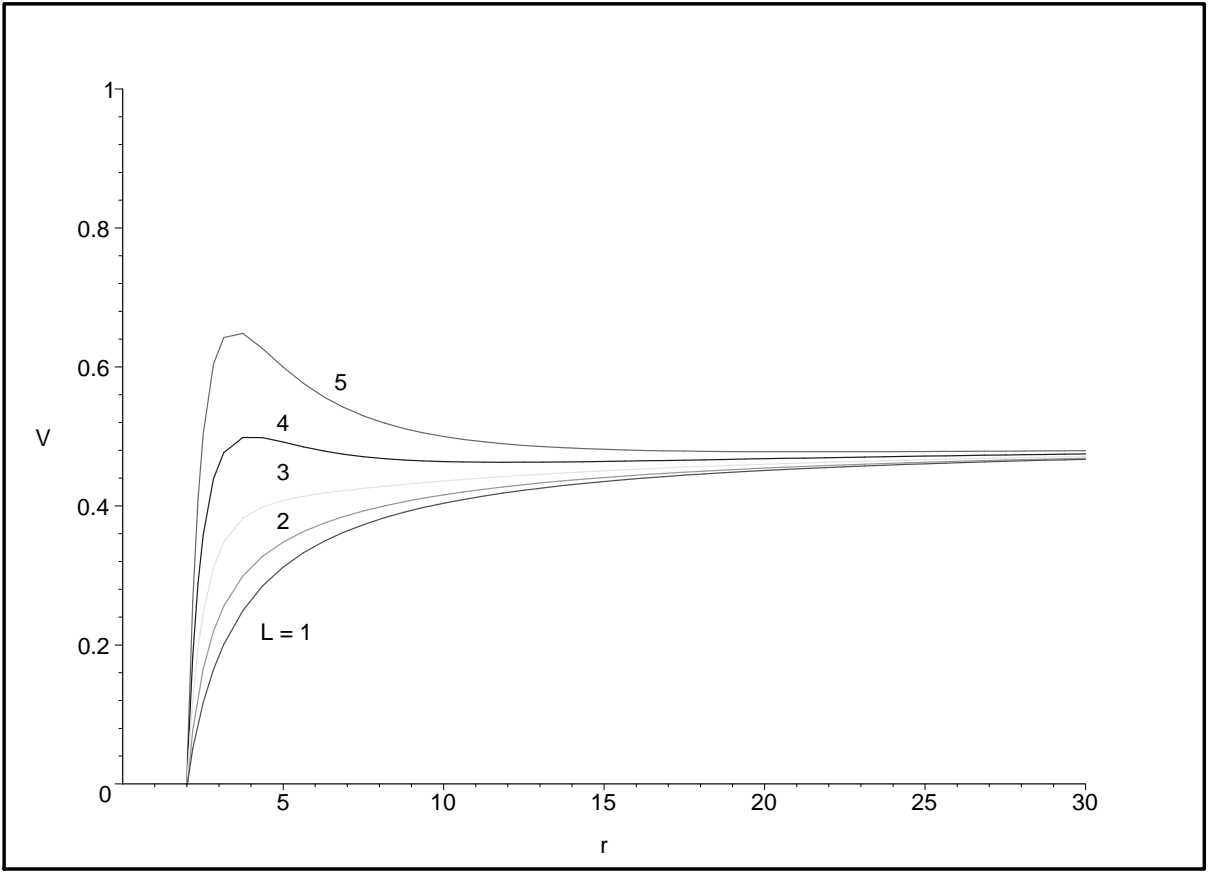}
\end{center}
\vspace{16 cm} \caption{\small {Commutative effective potential
versus radius for different values of angular momentum(in arbitrary
units).}}
 \label{fig:3}
\end{figure}

\begin{figure}[ht]
\begin{center}
\includegraphics{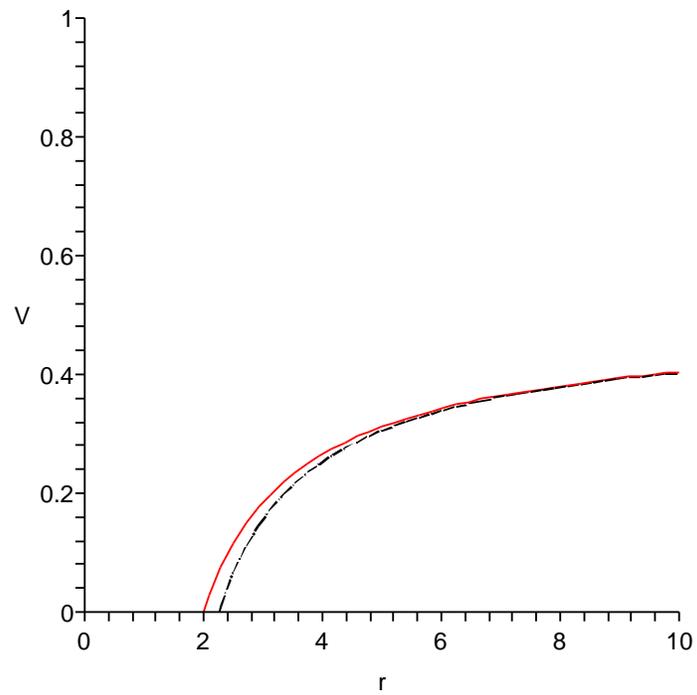}
\end{center}
\vspace{16 cm} \caption{\small {The difference between commutative
and noncommutative effective potential for $L=1$(in arbitrary
units). Lower curve shows noncommutative case. For small values of
$L$, the difference is not considerable. }}
 \label{fig:4}
\end{figure}

\begin{figure}[ht]
\begin{center}
\includegraphics{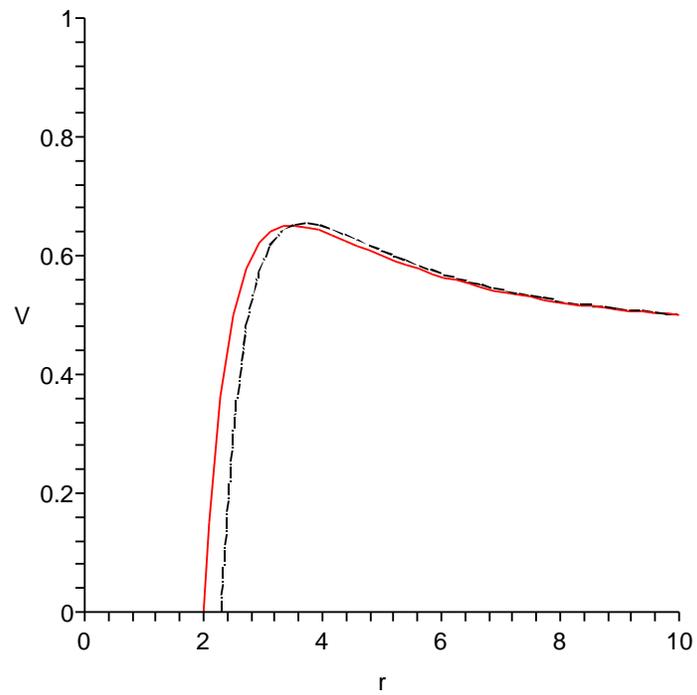}
\end{center}
\vspace{16 cm} \caption{\small {The difference between commutative
and noncommutative effective potential for $L=5$. Lower curve shows
noncommutative case. For large values of $L$, the difference is
considerable.}}
 \label{fig:5}
\end{figure}

\begin{figure}[ht]
\begin{center}
\includegraphics{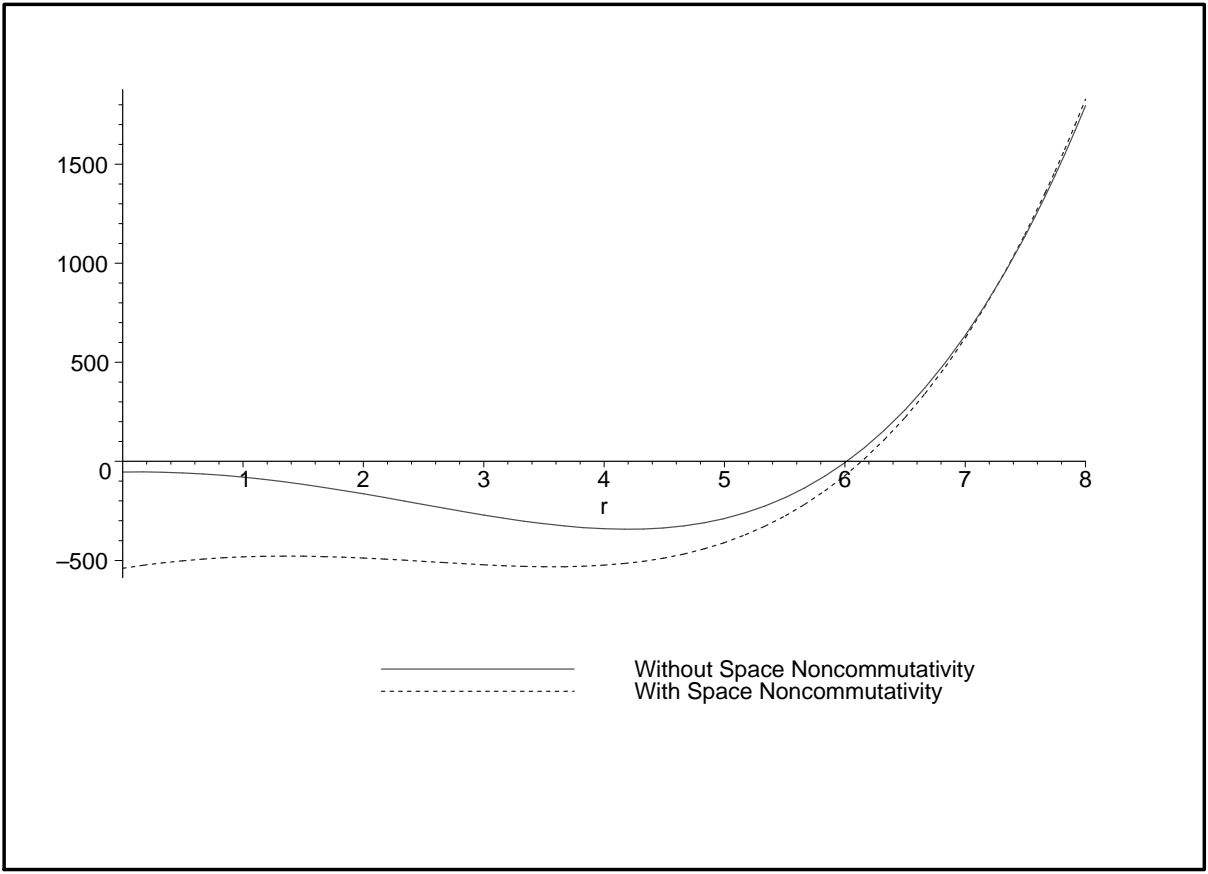}
\end{center}
\vspace{16 cm} \caption{\small {The condition for stability of
circular orbits for commutative and noncommutative Schwarzschild
spaces. In commutative case, particles could have stable circular
orbits for $r\geq6GM$ while for noncommutative case
$r\geq(6.139234690)GM$. Space noncommutativity increases the radius
of stable circular orbits.}}
 \label{fig:6}
\end{figure}

\end{document}